\documentclass[12pt]{article}
\usepackage[dvips]{graphicx}
\begin{document}
\begin{center}
{\Large \bf Spinor techniques for massive fermions
with arbitrary polarization}\\
\vspace{2cm}
{\large V. V. Andreev}\\
\smallskip
{\it Gomel State University, Physics Department,246699 Gomel, Belarus \\
and Abdus Salam International Centre For Theoretical Physics,\\
Trieste, Italy}\\
\end{center}


\begin{abstract}
We present a new variant of the spinor techniques for calculating
the amplitudes of processes involving  massive fermions
with arbitrary polarization. It is relatively simple
and leads to basic spinor products. Our procedure is
not more complex than CALCUL spinor techniques for massless
fermions. We obtain spinor
Chisholm identities for massive fermions.
As an illustration, expressions are given for the amplitudes of
electron-positron annihilation into fermions-pairs for several
polarizations.
\end{abstract}

\section {Introduction}

Studies of high-energy processes with polarization are of
fundamental importance in understanding structure of matter. For
example, The European Muon Collaboration (EMC) and SLAC experiments
on deep inelastic scattering of longitudinally polarized muons on
longitudinally polarized target have catalyzed an extraordinary
outburst of theoretical activity. With increasing energy of colliders,
processes involving many final-state
particles ( $2\to 3$, $2\to 4$, $2\to 5$, $\ldots$ ) are an
important part the present collider's physics.

The above-listed directions of high-energy physics have a point of
intersection. The calculation of cross sections for these processes
is difficult
if you used conventional approach.
This approach is to square the Feynman amplitude
and therefore it is very inconvenient to implement if the number of Feynman
diagrams and the number of final state particles are large.

An alternative approach is to compute the Feynman amplitudes
symbolically or numeri\-cally.
The idea of calculating Feynman amplitudes is as old
as the conventional approach. For example, covariant method of
calculating amplitudes have developed more thirty years ago
~\cite{sp2}-\cite{sp4}.

Many different methods are proposed for calculating the Feynman diagrams.
A particular spinor technique for massless external
fermions was introduced by the CALCUL collaboration ~\cite{sp5}-\cite{sp7}.
Generalizations of CALCUL approach to the massive fermion case exist
~\cite{sp6}-\cite{sp8}, but only for specific choice of the
spin projection. We call a this choice
{\bf KS-spin projection} or simple KS-states (these fermion states
often are called helicity states, but it is not correct).

A few methods of analytical calculations
of reactions with massive fermions are convenient for computer
symbolic calculations. Except the  above  mentioned  method  of  group
CALCUL, it is important to mention methods proposed in
Refs.~\cite{sp8},\cite{sp9}.  So in Ref.\cite{sp8} a compact
formalism of evaluating matrix elements based on the insertion
in spinor lines of a complete  set of states build up of unphysical
spinors was proposed, that has allowed  to create  a  high speed
program of evaluations of Feynman amplitudes.

The aim of this paper is to present a spinor techniques method
for calculating the amplitudes of processes involving massive
fermions with arbitrary polarizations. This approach lead to expressions
of Feynman amplitudes in terms of spinor products. As an illustration
we apply our method to compute amplitudes and cross sections of the
electron-positron annihilation into fermions-pairs for various usual
spin projection of fermion states (helicity, KS-states,
spin $z$-projection of the fermion in its rest frame).

\section{Spinor techniques for massless fermions}

In this section we briefly recall the spinor techniques of
Refs.\cite{sp5}-\cite{sp6}, but with small modifications.

Let us introduce the orthonormal four-vector basis  in Minkowski space
\footnote{These are covariant four vectors}
\begin{equation}
n_0=(1,0,0,0)~,n_1=(0,1,0,0),~n_2=(0,0,1,0),~n_3=(0,0,0,1).
\label{st1}
\end{equation}
They satisfy the completeness relation:
\begin{equation}
n_0^{\mu} \cdot n_0^{\nu} -n_1^{\mu} \cdot n_1^{\nu} -n_2^{\mu}
\cdot n_2^{\nu}-n_3^{\mu} \cdot n_3^{\nu} = g^{\mu \nu}
\label{st2}
\end{equation}
by means of which an arbitrary 4-vector $p$ can be written as:
\begin{equation}
p = p \cdot n_0 \cdot n_0 -p \cdot n_1 \cdot n_1 -p \cdot n_2 \cdot
n_2-p \cdot n_3 \cdot n_3 ~.  \label{st3} \end{equation}
Using (\ref{st1}) we can define light-like vectors \begin{equation}
b_0=n_0-n_3,b_3=n_0+n_3,b_\lambda =n_1+i\hskip 2pt \lambda n_2,
~~~\lambda=\pm 1
\label{st4}
\end{equation}
with following properties:
\begin{equation}
b_0 \cdot b_{\lambda}=0,~b_3 \cdot b_\lambda=0,~b_0 \cdot
b_3=2,~b_{+}\cdot b_{-}=-2, \label{st5} \end{equation}
\begin{equation} \frac{1}{2} \left ( b_0^{\mu} \cdot b_3^{\nu}
+b_3^{\mu} \cdot b_0^{\nu} -b_{+}^{\mu} \cdot
b_{-}^{\nu}-b_{-}^{\mu} \cdot b_{+}^{\nu} \right )= g^{\mu \nu}.
\label{st6} \end{equation}

Next we define {\it basic spinor } $U_\lambda \left( b_0 \right)$
by specifying the corresponding projection operator and phase condition:
\begin{equation}
U_\lambda \left( b_0\right) \overline{U}_\lambda \left( b_0\right) =\omega
_\lambda \not\!{b}_0,
\label{st7}
\end{equation}
\begin{equation}
\frac \lambda 2\not\!{b}_\lambda U_{-\lambda }\left( b_0\right) =U_\lambda
\left( b_0\right)
\label{st8}
\end{equation}
with matrix $\omega _{\lambda} = 1/2 \left( 1+\lambda
\gamma _5\right).$ Our $b_0$ corresponds to the vector called
$k_0=(1,1,0,0)$ in Ref.\cite{sp6} and
instead of $k_1=(0,0,1,0)$ we use $b_{\lambda}$.

The CALCUL spinor techniques for calculating processes with massless
external fermions involve the following operations:

{\large 1 step:}  The arbitrary massless spinor $U_\lambda \left( p \right)$
of momentum $p$ and helicity $\lambda$ is defined in terms of basic spinor
\begin{equation}
U_\lambda \left(p\right) = \frac{\not\!{p}}{
\sqrt{2 \hskip 1pt p \cdot b_0}}
\hskip 1pt U_{-\lambda }\left(b_0\right) ,
\label{st9}
\end{equation}
where $p \not = const \hskip 1pt b_0$.

{\large 2 step:} Using the spinor Chisholm identity
\begin{equation}
\gamma ^\mu \left\{ \overline{U}_\lambda \left( p\right) \gamma _\mu
U_\lambda \left( k\right) \right\}
=2\hskip 1pt U_\lambda \left( k\right) \overline{U}_\lambda \left(
p\right) +2 \hskip 1pt U_{-\lambda }\left( p\right)
\overline{U}_{-\lambda }\left( k\right)
\label{st10}
\end{equation}
and the equation for any real four-vector $p$ with $p^2=0$
\begin{equation}
\not\!{p}=\sum_\lambda U_\lambda \left( p\right)
\overline{U}_\lambda \left( p\right)
\label{st11}
\end{equation}
we can reduce the amplitudes of processes with massless fermions
to expressions involving  spinor products (or inner products)
\begin{equation}
s_{\lambda}\left(p,k\right) \equiv
\overline{U}_\lambda \left(p\right) U_{-\lambda}
\left( k\right)=-s_{\lambda }\left(k,p\right).
\label{st12}
\end{equation}
The remaining possible spinor products vanish due to
helicity conservation or are reduced to $s_{\lambda}$
with the help the of the relation
$$
V_\lambda \left( p\right) =U_{-\lambda }\left( p\right),
$$
where $V_\lambda \left( p\right)$ is the bispinor of an antifermion.

As an illustration we present the amplitudes for $e^+ e^-\to f
\bar f$, where $f$ is fermion ($f\not=e$). The Feynman diagrams
for this process are show in Fig.1. Using standard rules the
amplitude can be written as
\begin{figure}
\begin{center}
\resizebox{0.48\textwidth}{!}{
\includegraphics{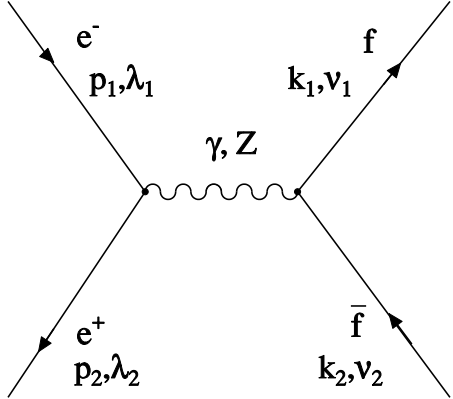}}
\end{center}
\caption{Feynman diagrams for the process $e^+ e^-\to f \bar f$}
\end{figure}

\begin{equation}
T\left (\lambda_1,\lambda_2;\nu_1,\nu_2 \right)=
\frac{4\pi\alpha}{s}
\left[ T_{\gamma} \left (\lambda_1,\lambda_2;\nu_1,\nu_2 \right)+
T_{Z^0} \left (\lambda_1,\lambda_2;\nu_1,\nu_2 \right) \right],
\label{st13}
\end{equation}
where
\begin{equation}
T_{\gamma} \left ( \lambda_1,\lambda_2;\nu_1,\nu_2 \right)
=Q_f \overline{V}_{\lambda_{2} } \left(p_2, s_{p_{2}} \right)
\gamma_{\mu} U_{\lambda_{1} } \left(p_1,s_{p_{1}}\right)
\overline{U}_{\nu_{1} } \left(k_1,s_{k_{1}}\right) \gamma^{\mu}
V_{\nu_{2} } \left( k_2, s_{k_{2}} \right),
\label{st14}
\end{equation}
\begin{eqnarray}
&&T_{Z^0} \left ( \lambda_1,\lambda_2;\nu_1,\nu_2 \right)=
R_z \left ( g^{\mu\nu}-q^\mu q^\nu/M_Z^2  \right )
\nonumber\\&&
\overline{V}_{ \lambda_{2} } \left( p_2, s_{p_{2}} \right)
\gamma_{\nu} \left( g_v^e-g_a^e \gamma_5 \right )
U_{\lambda_{1}} \left( p_1,s_{p_{1}} \right)
\overline{U}_{\nu_{1}} \left( k_1,s_{k_{1}} \right)
\gamma_{\mu} \left( g_v^f-g_a^f \gamma _5 \right )
V_{\nu_{2}}  \left( k_2, s_{k_{2}} \right),
\label{st15}
\end{eqnarray}
with $s=q^2=(p_1+p_2)^2 $, $ R_z=\left( G_F M_Z^2 s \right)/
\left(2 \sqrt{2}\pi \alpha \left(s-M_Z^2 \right) \right)$.
The $g_v^f, g_a^f$  are fermion couplings and $\alpha=e^2/(4\pi)$, $G_F$ are
the fine-structure and the Fermi constant respectively, $Q_f$ is
the $f$ charge in units $e$ .
The notation $s_p$ for
bispinors indicates that fermion with momentum $p$ has fixed
polarization vector $s_p$.
Using (\ref{st10}), the nonzero amplitude in {\bf massless case} can be
written in terms of the spinor products (\ref{st12}):
\begin{eqnarray}
&&T \left ( \lambda,-\lambda;\nu,-\nu \right )=
\frac{8\pi\alpha}{s}
\left( Q_f+R_z \left ( g_v^e-\lambda\hskip 1pt g_a^e \right )
\left ( g_v^f-\nu \hskip 1pt g_a^f \right ) \right )
\nonumber\\&&
\left [
\delta_{\lambda,\nu} s_{\lambda}\left(k_1,p_2\right)
s_{-\lambda} \left(p_1,k_2 \right)+
\delta_{\lambda,-\nu} s_{\lambda} \left(p_2,k_2\right)
s_{-\lambda} \left(k_1,p_1\right)  \right ] ~.
\label{st16}
\end{eqnarray}
Thus, in the evaluation of the
Feynman diagrams with the help of spinor techniques
the spinor products are important 'building' blocks.

The spinor product (\ref{st12})
due to equations (\ref{st7})-(\ref{st8}) reduces to the trace
\footnote{
$\epsilon\left (b_0\hskip 1pt, b_\lambda \hskip 1pt,
p \hskip 1pt, k \right )=\epsilon _{\mu \nu \rho \sigma
}
b_0^\mu \hskip 1pt b_\lambda ^\nu \hskip 1pt p^\rho \hskip 1pt k^\sigma$,
$\epsilon ^{0\hskip 1pt 1\hskip 1pt 2\hskip 1pt3}=-1$ }
\begin{eqnarray}
&&s_\lambda \left(p,k\right) =
\frac{\lambda}{4}\frac{ \hskip 2pt
Tr\left(\omega _{-\lambda }\not\!{b}_0 \hskip 1pt
\not\!{p} \hskip 1pt \not\!{k}
\hskip 1pt \not\!{b}_\lambda \right) }
{\sqrt{b_0 \cdot p}\sqrt{b_0 \cdot k}} =
\\
\label{st17}
&& \frac{\lambda}{2} \frac{
\left[p.b_0\hskip 2pt k .b_\lambda -k.b_0 \hskip 2pt p.b_\lambda\right] -
i\epsilon\left (b_0\hskip 1pt, b_\lambda \hskip 1pt,
p \hskip 1pt, k \right )}
{\sqrt{b_0 \cdot p}\sqrt{b_0 \cdot k} } =
\\
\label{st18}
&&\frac{\lambda
\left[ p\cdot b_0\hskip 2pt k \cdot b_\lambda -k \cdot b_0 \hskip 2pt
p \cdot b_\lambda \right]}{\sqrt{b_0 \cdot p} \sqrt{b_0 \cdot k}}.
\label{st19}
\end{eqnarray}
From Eqs.(\ref{st4}),(\ref{st6}) for a real light-like vector $p$
it follows that
\begin{equation}
\left |p \cdot b_{+}\right|=\left |p \cdot b_{-}
\right|=\sqrt{p \cdot b_0~ p \cdot b_3}~~.  \label{st20} \end{equation}
If we
define \begin{equation} p \cdot b_0=p^0-p^z\equiv p^-,\qquad
p\cdot b_3=p^0+p^z\equiv p^+, p \cdot b_{\lambda}=\sqrt{p^+~p^-}\exp \left( i
\lambda \varphi _p\right), \label{st21} \end{equation}
then it
follows that \begin{equation} s_{\lambda}\left(p,k\right) =
\lambda \left ( \sqrt{p^-~k^+} \exp \left( i \lambda \varphi _k\right)-
\sqrt{p^+~k^-} \exp \left( i \lambda \varphi _p\right)
\right ).
\label{st22}
\end{equation}
The spinor product (\ref{st22}) of two four-vectors is not
more complicated than the dot Minkowski product $p.k$.
It is important, that definition of the spinor product (\ref{st22})
is valid for any real 4-vectors, including also $b_0$ and $b_3$.

There are several possibilities to calculate the amplitude (\ref{st13}).
We can construct an orthonormal basis (\ref{st1})
with the help of the physical vectors $p_1,p_2,k$
\begin{equation}
n_0=\frac{p_1+p_2 }{\sqrt{2\hskip 1pt p_1 \cdot p_2} },
\qquad
n_3=\frac{p_2-p_1 }{\sqrt{2\hskip 1pt p_1 \cdot p_2} },
\qquad
n_1^{\mu}=\epsilon (\mu,n_0,n_3,n_2),
\qquad
n_2^{\mu}= \frac{\epsilon\left( \mu,p_1,p_2,k_1 \right) }
{\sqrt{2 p_1 \cdot p_2 \hskip 1pt p_1 \cdot k_1 \hskip 1pt p_2 \cdot k_1}  }
\label{st23}
\end{equation}
and obtain an analytic expression for the amplitude in terms of dot products.

Also, using (\ref{st22}), we can write the amplitude in terms of
the components of the momenta. Finally, we have the possibility
of calculating numerically the amplitude with the help of
Eqs.(\ref{st16}), (\ref{st22}).

\section {Spinor techniques for massive fermions}

Our procedure will be similar to that presented in the above section.
Let us consider bispinors, which are related to the basic spinor
(see appendix of Ref.\cite{sp7}) by
\begin{equation}
U_\lambda \left( p,s_p \right) =\frac{\tau _u^\lambda \left( p,s_p \right) }
{\sqrt{ b_0 \cdot \left( p+m_p s_p\right) }}U_{-\lambda }\left( b_0\right) ,
\label{st24}
\end{equation}
\begin{equation}
V_\lambda \left( p,s_p\right) =\frac{ \tau _v^\lambda \left( p,s_p\right) }
{\sqrt{ b_0 \cdot \left( p+m_p s_p\right) }}U_\lambda \left( b_0\right),
\label{st25}
\end{equation}
where the projection operators $\tau
_u^\lambda \left( p,s_p\right), \tau _v^\lambda \left( p,s_p\right)$
are
\begin{equation}
\tau _u^\lambda \left( p,s_p\right) =\frac
1{2}\left( \not\!{p}+m_p\right) \left( 1+\lambda \hskip 2pt \gamma
_5\hskip 2pt \not\!{s_p}\right),
\label{st26}
\end{equation}
\begin{equation}
\tau _v^\lambda \left( p,s_p\right) =\frac 1{2}\left( \not\!{p}-m_p\right)
\left( 1+\lambda \gamma _5\hskip 2pt \not\!{s_p}\right).
\label{st27}
\end{equation}
We obtain
\begin{eqnarray}
&&\not\!{p} \hskip 2pt U_\lambda \left( p,s_p\right) = m_p \hskip 2pt
U_{\lambda }\left( p,s_p\right),
\hskip 7pt
\not\!{p} \hskip 2pt V_\lambda \left(
p,s_p\right) = - m_p \hskip 2pt V_{\lambda }\left( p,s_p\right) ,
\nonumber\\&&
\gamma_5 \not\!{s_p} \hskip 2pt U_\lambda \left( p,s_p\right) =
\hskip 7pt
\lambda \hskip 2pt U_{\lambda }\left( p,s_p\right),
\gamma_5\not\!{s_p} \hskip 2pt V_\lambda \left( p,s_p\right) = \lambda
\hskip 2pt V_{\lambda }\left( p,s_p\right)
\label{st28}
\end{eqnarray}
i.e. the
bispinors $U_\lambda \left( p,s_p\right)$ and $V_\lambda \left(
p,s_p\right) $ satisfy Dirac equation and spin condition for
massive fermion and antifermion. We also found, that the bispinors of
fermion and antifermion (\ref{st24})-(\ref{st25}) are related by
\begin{equation}
V_\lambda \left( p,s_p\right)=
-\lambda \gamma _5U_{-\lambda }\left( p,s_p\right),
\hskip 7pt
\overline{V}_\lambda \left( p,s_p\right) =\overline{U}_{-\lambda }\left(
p,s_p\right) \lambda \hskip 2pt \gamma_5.
\label{st29}
\end{equation}

Let us analyze how many  spinor products there are for massive
fermions.  Obviously, the case of massive fermions is more difficult
in comparison with the massless one. In the general case for calculating
the matrix elements  sixteen spinor products are necessary while in
the massless case there are only two. However,
it is possible to achieve essential simplification.

We notice that the  spinor products are not all linearly independent.
Using relations
(\ref{st29}) we can show, that only eight products are linearly independent.
We define these {\it basic spinor products } for massive fermions as
\begin{equation}
\overline{U}_\lambda \left( p,s_p\right) U_{-\lambda }\left( k,s_k\right) ,
\qquad
\overline{U}_\lambda \left( p,s_p\right) U_\lambda \left( k,s_k\right) ,
\qquad
\overline{V}_\lambda \left( p,s_p\right) U_\lambda \left( k,s_k\right) ,
\qquad
\overline{V}_\lambda \left( p,s_p\right) U_{-\lambda }\left( k,s_k\right) .
\label{st30}
\end{equation}
As one can see below, the basic spinor products (\ref{st30})
can be calculated with the help of two functions.

The matrix $\gamma _5$  is the
operator of a spin for massless bispinor i.e.
\begin{equation}
\gamma _5 U_\lambda \left( b_0\right) =\lambda U_\lambda \left( b_0\right) .
\label{st31}
\end{equation}
Then the definition of bispinors (\ref{st24})-(\ref{st25})
can be rewritten as
\begin{equation}
U_\lambda \left( p,s_p \right) = \frac{\chi\left( p,s_p,+1 \right) }{2 \sqrt{
b_0 \cdot \left( p+m_p s_p\right) }}U_{-\lambda }\left( b_0\right) ,
\label{st32}
\end{equation}
\begin{equation}
V_\lambda \left(
p,s_p\right) =\frac{ \chi \left( p,s_p,-1 \right) } {2 \sqrt{
b_0 \cdot \left( p+m_p s_p\right) }}U_\lambda \left(b_0\right),
\label{st33}
\end{equation}
with the function
\begin{equation}
\chi\left( p,s_p,a \right) = \left( \not\!{p}+a \hskip 1pt m_p \right)
\left( 1+a \hskip 1pt \not\!{s_p} \right).
\label{st34}
\end{equation}

Let us introduce the following functions:
\begin{equation}
\upsilon\left( p,k,s_p,s_k,\lambda,a\right) \equiv
\frac{\lambda}{8}
\frac{
Tr \left( \omega_{-\lambda }\hskip 1pt \not\!{b}_0 \hskip 1pt
\chi^{\dagger} \left( p,s_p,a \right) \chi \left( k,s_k,+1 \right)
\hskip 1pt \not\!{b}_\lambda \right)}{
\sqrt{ b_0 \cdot  \left( p+m_p \hskip 2pt s_p \right)} \sqrt{ b_0 \cdot  \left(
k+m_k \hskip 2pt s_k \right) }},
\label{st35}
\end{equation}
\begin{equation}
w\left(p,k,s_p,s_k,\lambda,a \right) \equiv
\frac{1}{4}
\frac{
Tr \left(
\omega_{-\lambda } \hskip 1pt \not\!{b}_0 \hskip 1pt \chi^{\dagger}
\left( p,s_p,a \right) \chi \left( k,s_k,+1 \right) \right)}
{ \sqrt{ b_0 \cdot
\left( p+m_p \hskip 2pt s_p \right) } \sqrt{ b_0 \cdot  \left( k+m_k \hskip
2pt s_k \right) }} .
\label{st36}
\end{equation}
All the basic spinor products are reduced then to functions
(\ref{st35})--(\ref{st36}):
\begin{eqnarray}
\overline{U}_\lambda \left( p,s_p\right) U_{-\lambda }\left(
k,s_k\right)
&=& \upsilon\left(p,k,s_p,s_k,\lambda,a=+1\right),
\nonumber
\\
\overline{U}_{\lambda}\left( p,s_p\right) U_\lambda \left(
k,s_k\right)
&=& w\left(p,k,s_p,s_k,\lambda,a=+1 \right),
\nonumber
\\
\overline{ V}_{\lambda} \left(p,s_p\right) U_\lambda \left(
k,s_k\right)
&=&\upsilon\left(p,k,s_p,s_k,-\lambda,a=-1\right),
\nonumber
\\
\overline{V}_{\lambda} \left( p,s_p\right) U_{-\lambda}
\left(k,s_k\right)
&=&
w\left(p,k,s_p,s_k,-\lambda,a=-1\right).
\label{st37}
\end{eqnarray}
As for the massless fermions, the spinor products (\ref{st37}) can be
calculated through the components of vectors $p,k,s_p,s_k$.
Certainly, the analytical expressions (\ref{st37}) are more
complex on a comparision with the appropriate formulas
(\ref{st19}),(\ref{st22}).

An important role in the transformation  of the matrix  elements  to  basis
spinor products is played by the spinor identities Chisholm  of type
(\ref{st10}). For a proof of relation (\ref{st10})
the Chisholm trace identity was used in Ref.\cite{sp6}:
\begin{equation}
\gamma^{\mu}\hskip 2pt Tr(\gamma_{\mu} \not\!{C}_1\ldots
\not\!{C}_{2n+1}) =2 \left( \not\!{C}_1 \ldots \not\!{C}_{2n+1} +
\not\!{C}_{2n+1}\ldots \not\!{C}_1\right) .  \label{st38}
\end{equation}

As an expression of the type $ \overline{U}_\lambda  \left(p,s_p\right)
\gamma_\mu U_\lambda \left (k,s_k\right)$ is reduced
to trace, using the formulas (\ref{st24})-(\ref{st25}), (\ref{st38}) it
is possible to obtain appropriate  spinor  identities  for  massive
fermions. In this case we have four basic identities
(compared to one in the massless case):
\begin{eqnarray}
&&\gamma ^\mu \left\{
\overline{U}_\lambda \left( p,s_p\right) \gamma _\mu U_\lambda
\left( k,s_k\right) \right\}
=U_\lambda \left( k,s_k\right)
\overline{U}_\lambda \left( p,s_p\right) +U_{-\lambda }\left(
p,s_p\right) \overline{U}_{-\lambda }\left( k,s_k\right) +
\nonumber\\&&
+V_{-\lambda }\left( k,s_k\right) \overline{V}_{-\lambda }\left(
p,s_p\right) +V_\lambda \left( p,s_p\right) \overline{V}_\lambda \left(
k,s_k\right) ,
\label{st39}
\end{eqnarray}
\begin{eqnarray}
&&\gamma ^\mu \left\{ \overline{U}_\lambda \left( p,s_p\right)
\gamma _\mu U_{-\lambda }\left( k,s_k\right) \right\}
=U_{-\lambda }\left( k,s_k\right) \overline{U}_\lambda \left( p,s_p\right)
-U_{-\lambda }\left( p,s_p\right) \overline{U}_\lambda \left( k,s_k\right) +
\nonumber\\&&
+V_\lambda \left( p,s_p\right) \overline{V}_{-\lambda }\left( k,s_k\right)
-V_\lambda \left( k,s_k\right) \overline{V}_{-\lambda }\left( p,s_p\right) ,
\label{st40}
\end{eqnarray}
\begin{eqnarray}
&&\gamma ^\mu \left\{ \overline{V}_\lambda \left( p,s_p\right)
\gamma _\mu U_{-\lambda }\left( k,s_k\right) \right\}
=U_{-\lambda }\left( k,s_k\right) \overline{V}_\lambda \left(
p,s_p\right) +V_{-\lambda }\left( p,s_p\right) \overline{U}_\lambda
\left( k,s_k\right)+
\nonumber\\&&
+V_\lambda \left(
k,s_k\right) \overline{U}_{-\lambda }\left( p,s_p\right) +U_\lambda
\left( p,s_p\right) \overline{V}_{-\lambda }\left( k,s_k\right) ,
\label{st41}
\end{eqnarray}
\begin{eqnarray}
&&
\gamma ^\mu \left\{ \overline{V}_\lambda \left( p,s_p\right) \gamma
_\mu U_\lambda \left( k,s_k\right) \right\} =U_\lambda \left(
k,s_k\right) \overline{V}_\lambda \left( p,s_p\right) +U_\lambda
\left( p,s_p\right) \overline{V}_\lambda \left( k,s_k\right)-
\nonumber\\&&
-V_{-\lambda }\left( k,s_k\right) \overline{U}_{-\lambda }\left(
p,s_p\right) -V_{-\lambda }\left( p,s_p\right) \overline{U}_{-\lambda
}\left( k,s_k\right) .
\label{st42}
\end{eqnarray}
The remaining possible combinations are reduced to the above ones  with
the help of relations (\ref{st29}). For massless fermions the relations
(\ref{st39}) and  (\ref{st41})  are identical (to  within the
replacement $\lambda \to -\lambda$) and pass into (\ref{st10}),
and Eqs.(\ref{st40}) and (\ref{st42}) both sides are zero.

Using (\ref {st39})-(\ref{st42}),  and Dirac equation
we  can write the analytical expression of a matrix element (\ref{st13})
in terms of spinor products. We represent the amplitude by
$T\left ( \lambda,-\lambda; \lambda,-\lambda \right)$
for {\bf massive fermions with arbitrary
polarizations} as an example of our type of spinor techniques:
\begin{eqnarray}
&&
T\left(\lambda, -\lambda; \lambda, -\lambda  \right)
=\frac{4\pi\alpha}{s}
\left \{  \right.
\left(Q_f+R_z\left(g_a^e g_a^f+g_v^e g_v^f \right)\right)
\nonumber
\\
&&
\left[
\upsilon\left(k_1, p_2, \lambda, -1 \right)
\upsilon\left(p_1, k_2,-\lambda, -1 \right) +
\upsilon\left(k_1, p_2, \lambda,  1 \right)
\upsilon\left(p_1, k_2,-\lambda,  1 \right)
\right] -
\nonumber
\\
&&
\lambda R_z\hskip 1pt \left(g_a^f g_v^e+g_a^e g_v^f \right )
\left[
\upsilon\left(k_1, p_2, \lambda, 1 \right)
\upsilon\left(p_1, k_2,-\lambda,-1 \right) +
\upsilon\left(k_1, p_2, \lambda,-1 \right)
\upsilon\left(p_1, k_2, -\lambda,1 \right)
\right] -
\nonumber
\\
&&
\left(Q_f+ R_z \left(g_a^e g_a^f-g_v^e g_v^f \right)\right)
\left[
w\left(k_1, p_1, \lambda, -1\right)
w\left(p_2, k_2, \lambda, -1\right) -
w\left(k_1, p_1, \lambda,  1\right)
w\left(p_2, k_2, \lambda,  1\right)
\right] -
\nonumber
\\
&&
\lambda R_z \hskip 1pt \left(g_a^f g_v^e - g_a^e g_v^f \right)
\left[
w\left(k_1, p_1, \lambda,  1\right)
w\left(p_2, k_2, \lambda, -1\right) -
w\left(k_1, p_1, \lambda, -1\right)
w\left(p_2, k_2, \lambda,  1\right)
\right] -
\nonumber
\\
&&
4 R_z g_a^e g_a^f m_k m_p
w\left(k_1, k_2, \lambda, 1\right)
w\left(p_2, p_1, \lambda, 1\right)/M_Z^2
\left. \right \} ,
\label{st43}
\end{eqnarray}
where we use the notation $ \upsilon \left( p_1, k_2, \lambda,1
\right) \equiv \upsilon \left( p,k,s_p,s_k, \lambda, 1 \right)$
(and a similar one for $w$). Analogous expressions can be written
for other spin configurations of fermions. The amplitude with
massive fermions looks more complicated than the massless one
(\ref{st16}),but we will see below, that the amplitude
(\ref{st43}) has a simple form.

\section {Spinor products}

Let us calculate the functions $\upsilon\left(p,k,s_p,s_k, \lambda,a\right)$
and $w\left(p,k,s_p,s_k, \lambda,a\right)$ for several
spin projection of the fermion states.

The polarization vector $s_p$  of  a  fermion  can  be  expressed
through the momentum of the fermion as
\begin{equation}
s_p=\frac{p \cdot q_p \hskip 2pt p -m^2_p \hskip 2pt q_p}
{m_p\sqrt{(p \cdot q_p)^2-m^2_p \hskip 2pt q_p^2}},
\label{st44}
\end{equation}
where  $q_p$ is an arbitrary vector ($q_p\not = const~p $).
It is easy see, that $s_p$ satisfies to standard conditions:
\begin{equation}
s_p^2=-1,~~p \cdot s_p=0.
\label{st45}
\end{equation}
Choosing
\begin{equation}
q_p = n_0 = (1,0,0,0)
\label{st46}
\end{equation}
we find in this case that the state of polarization of a fermion
is the {\bf helicity state }.

Taking
\begin{equation}
q_p=b_0=n_0-n_3
\label{st47}
\end{equation}
then the polarization vector (\ref{st44}) can be written as
\begin{equation}
s_p = \frac{p}{m_p}-m_p \hskip 2pt \frac{b_0}{p \cdot b_0} .
\label{st48}
\end{equation}

Let us call the fermion states with this choice of the spin quantization
vector the {\bf KS states}(see \cite{sp6}).
As we see, helicity states and KS states are
different in the general case.

Finally, if in the rest frame of the fermion $s_p=n_3$
(axis of spin projection is $z$-axis) we have the fermion polarized states,
which will be called the {\bf $z$ states}.

The spinor products can be rewritten as functions
\begin{eqnarray}
&&\upsilon\left(p,k,s_p,s_k, \lambda,a\right)\equiv
\upsilon\left(p,k,q_p,q_k, \lambda,a\right),
\nonumber
\\
&&
w\left(p,k,s_p,s_k, \lambda,a\right)\equiv
w\left(p,k,q_p,q_k, \lambda,a\right).
\label{st49}
\end{eqnarray}

For the calculation of the amplitudes it is often assumed that
$q_p=q_k=\ldots \equiv q$  i.e. all fermions have
the same polarized states (helicity, KS, $z$ or another possible state).
The spinor products $\upsilon$ and $w$ are denoted
\begin{equation}
\upsilon_{hel}\left(p,k,\lambda,a\right),
w_{hel}\left(p,k,\lambda,a\right),
\hskip 7pt
\upsilon_{KS}\left(p,k,\lambda,a\right),
w_{KS}\left(p,k,\lambda,a\right),
\hskip 7pt
\upsilon_{z}\left(p,k,\lambda,a\right),
w_{z}\left(p,k,\lambda,a\right)
\label{st50}
\end{equation}
for helicity, KS and $z$ states respectively.

Using (\ref{st35})-(\ref{st36}) we express the functions
$\upsilon$ and $w$ through the components of the vectors.
For the vectors $p =\left(p^0,p^x,p^y,p^z\right)$ and $k
=\left(k^0,k^x,k^y,k^z\right)$ we obtain:
\begin{eqnarray}
&&
\upsilon_{hel}\left( p,k,\lambda,a \right)=
\frac{\lambda
\left ( a\hskip 1pt m_p\hskip 1pt m_k+
\left( k^0+ \left| \vec {k}\right| \right)
\left ( p^0+\left|\vec {p}\right| \right ) \right)}
{\sqrt{4\left|\vec {k}\right|\left|\vec {p}\right|
\left(k^0+\left|\vec {k}\right|\right)
\left(p^0+\left|\vec {p}\right|\right)}}
\nonumber \\ &&
\left \{
e^{i\lambda \varphi_k} \sqrt{ \left(\left|\vec {k}\right|+k^z\right)
\left(\left|\vec {p}\right|-p^z\right)}
-e^{i\lambda \varphi_p} \sqrt{\left( \left|\vec {k}\right|-k^z\right)
\left(\left|\vec {p}\right|+p^z\right)}
\right \},
\label{st51}
\end{eqnarray}
\begin{eqnarray}
&&
w_{hel}\left( p,k,\lambda,a \right)=
\frac{
\left ( a \hskip 1pt m_p \left( k^0+ \left| \vec {k}\right| \right)+
m_k \left ( p^0+\left|\vec {p}\right| \right )
\right)}{
\sqrt{4\left|\vec {k}\right|\left|\vec {p}\right|
\left(k^0+\left|\vec {k}\right|\right)
\left(p^0+\left|\vec {p}\right|\right)}}
\nonumber \\ &&
\left \{
\sqrt{ \left(\left|\vec {k}\right|-k^z\right)
\left(\left|\vec {p}\right|-p^z\right)}
+e^{i\lambda \left(\varphi_k-\varphi_p\right)}
\sqrt{\left( \left|\vec {k}\right|+k^z\right)
\left(\left|\vec {p}\right|+p^z\right)}
\right \},
\label{st52}
\end{eqnarray}
\begin{eqnarray}
&&\upsilon_{KS}\left( p,k,\lambda,a\right) =
\lambda
\left\{
\right.
e^{i\lambda \varphi_k}
\sqrt{p^0-p^z} \sqrt{ \left( k^0+k^z \right)-m_k^2/\left( k^0-k^z \right)}-
e^{i\lambda \varphi_k}
\nonumber \\ &&
\sqrt{k^0-k^z} \sqrt{ \left( p^0+p^z \right)-m_p^2/\left( p^0-p^z \right)}
\left.
\right\},
\label{st53}
\end{eqnarray}
\begin{equation}
w_{KS} \left( p,k,\lambda ,a\right)=
a\hskip 1pt m_p
\hskip 2pt \sqrt{\left(k^0-k^z\right)/\left (p^0-p^z\right )}+
m_k
\hskip 2pt \sqrt{\left(p^0-p^z\right)/\left (k^0-k^z\right )},
\label{st54}
\end{equation}
\begin{eqnarray}
&&\upsilon_{z}\left( p,k,\lambda,a \right)=
\frac{
\lambda}{ \sqrt{4\left(k^0+m_k\right)
\left(p^0+m_p\right)}}
\nonumber \\ &&
\left \{
\right.
e^{i\lambda \varphi_p}
\sqrt{{\left|\vec {p}\right|}^2-(p^z)^2}
\left [
\left ( a-1 \right )\left ( k^0+m_k  \right )+
\left ( a+1  \right )k^z
\right ]
\nonumber \\ &&
-e^{i\lambda \varphi_k}
\sqrt{{\left|\vec {k}\right|}^2-(k^z)^2}
\left [
\left ( a-1 \right )\left ( p^0+m_p  \right )+
\left ( a+1  \right ) p^z
\right ]
\left.
\right\},
\label{st55}
\end{eqnarray}
\begin{eqnarray}
&& w_{z}\left( p,k,\lambda,a \right)=
\frac{1}{\sqrt{4 \left(k^0+m_k\right)
\left(p^0+m_p\right)}}
\nonumber \\ &&
\left \{
\right.
\left [
\left ( a+1 \right )\left ( k^0+m_k  \right )+
\left (1-a  \right )k^z
\right] \left ( p^0+m_p  \right)
+
\left[ \right.
\left (a-1 \right )\left ( k^0+m_k  \right )-
\nonumber \\ &&
\left (1+a  \right ) k^z
\left. \right ] p^z-
\lambda e^{i\lambda\left( \varphi_p-\varphi_k\right)}
\sqrt{\left( {\left|\vec {k}\right|}^2-(k^z)^2\right)
\left( {\left|\vec {p}\right|}^2-(p^z)^2\right)}
\left ( a+1  \right )
\left.
\right\}.
\label{st56}
\end{eqnarray}
As one can see, the analytical expressions
(\ref{st51})-(\ref{st56}) looks a little bit more complicated
in the corresponding relation (\ref{st22}).
In the massless limit,  the mass term in (\ref{st51})-(\ref{st54})
can be dropped. As a result, for the functions $\upsilon_{hel}$ and
$\upsilon_{KS}$ we obtain the expression (\ref{st22}).
Since all $z$ states have only one spin projection ($z$-axis in rest frame),
the massless limit is absent for these polarized states
in general.

Let us consider possible unstable situations, which can arise in numerical
calculations. There are the denominators in (\ref{st51})-(\ref{st56}),
hence the ambiguity of type $0/0$ can appear in the calculations.
Obviously we have not problems with helicity for massive and
massless fermions and  $z$, KS states for massive fermions.
If one chooses
the vector $k=(k^0,0,0,k^0) $ (the massless fermion moving along $z$-axis),
then the expressions for the KS states (\ref{st53})-(\ref{st54})
contain the ambiguity of type $0/0$.
In this situation for $\upsilon_{KS}$ and $w_{KS}$
we have the rule in the massless limit :\\
{\it 1 step.} To take the mass of the fermion equal to zero.\\
{\it 2 step.} Only after {\it 1 step} we calculate the spinor products
through the components of the vectors.

If another spin projection $s_k$ is interest of
we can decompose the any bispinor $U_\lambda \left( p,s_p \right)$
in terms of the another bispinor $U_\nu \left( k,s_k \right)$
with the help of spinor products, using completeness relation of the
bispinors (\ref{st24})- (\ref{st25}):
\begin{equation}
\sum\limits_{\nu}\frac{
U_\nu \left( k,s_k \right)
\overline{U}_\nu \left( k,s_k \right)-
V_\nu \left( k,s_k \right)
\overline{V}_\nu \left( k,s_k \right)
}{2 \hskip 1pt m_k}=1.
\label{st58}
\end{equation}

As an example we present  the matrix of decomposition
for KS and helicity states through the components of the momenta:
\begin{equation}
\frac{
\overline{U}_\lambda \left( p, hel \right)
U_\nu \left( p, KS \right)}{2 m_p}=
\delta_{\lambda,\nu}
\sqrt{
\frac{
\left(p^0+\left|\vec {p}\right|\right)
\left(\left|\vec {p}\right|-p^z\right)}{
2 \left|\vec {p}\right|\left(p^0-p^z\right)}} -
\lambda\delta_{\lambda,-\nu}
\hskip 1pt e^{i\lambda \varphi_p}
\sqrt{
\frac{\left(p^0-\left|\vec {p}\right|\right)
\left(\left|\vec {p}\right|+p^z\right)}{
2\left|\vec {p}\right|\left(p^0-p^z\right)
}}.
\label{st59}
\end{equation}

The other important building block is the element of the
current $J^{\mu}$ defined as follows
\begin{eqnarray}
&&J^{\mu}_u\left ( p,k,s_p ,s_k ;\lambda,\nu  \right )\equiv
\overline{U}_\lambda \left( p,s_p \right)
\gamma^{\mu} U_\nu \left( k,s_k \right),
\nonumber \\ &&
J^{\mu}_v\left ( p,k,s_p ,s_k; \lambda,\nu  \right )\equiv
\overline{V}_\lambda \left( p,s_p \right)
\gamma_{\mu} U_\nu \left( k,s_k \right).
\label{st60}
\end{eqnarray}
Using the completeness identity (\ref{st6}) we obtain
that the current $J$ of type (\ref{st60}) can be written
\begin{equation}
J^{\mu}\left ( p,k,s_p,s_k;\lambda,\nu  \right )=
\frac{1}{2} \left(  b_0 \cdot J \cdot b_3^{\mu} +
b_3 \cdot J \cdot b_0^{\mu} -b_{+}\cdot J \cdot
b_{-}^{\mu}-b_{-}\cdot J \cdot b_{+}^{\mu}
\right)
\label{st61}
\end{equation}
It is easy to verify that the terms $b_0 \cdot J$ and other
are reduced then to spinor products $\upsilon$ and $w$.
Using the analytical expressions (\ref{st51})-(\ref{st56})
we obtain coefficients of decomposition (\ref{st61})
through the components of the vectors (see these coefficients
for helicity states in appendix A).

We now have all the tools necessary to express any Feynman diagrams
with arbitrary fermion polarizations in terms of the spinor products
or the components of the momenta.

\section {Application}

As an illustration of our method we present the helicity and KS amplitudes
for process $e^+e^-\to f \bar f$.
Should be noted, that the spinor techniques with functions
(\ref{st53})-(\ref{st54})(KS states) are used for the
calculating of amplitudes of
$e^+e^-\to f \bar f+n \gamma$ with massless electron and positron
\cite{sp10}. Also we take the massless $e^+$ and $e^-$.
For completeness we have included expressions
of helicity amplitude (\ref{st13}) with all massive fermion
in appendix B.

We will work in the center of momentum system with
the initial particles moving along $z$-axis. In this frame the
momenta take the form
\begin{eqnarray}
&&p_1=\frac{\sqrt{s}}{2} (1,0,0,1)~ - ~~for~~ e^-,
\nonumber \\ &&
p_2=\frac{\sqrt{s}}{2} (1,0,0,-1)~ - ~~for~~ e^+,
\nonumber \\ &&
k_1=\frac{\sqrt{s}}{2} (1,
\beta_k \hskip 1pt sin\hskip 1 pt \theta, 0,
\beta_k \hskip 1pt cos \hskip 1pt \theta )~ - ~~for ~~f,
\nonumber \\ &&
k_2=\frac{\sqrt{s}}{2} (1,-\beta_k \hskip 1pt sin\hskip 1 pt \theta,0,
-\beta_k \hskip 1pt cos \hskip 1pt \theta)~ - ~~for~~ \bar f,
\label{st62}
\end{eqnarray}
where $s=(p_1+p_2)^2$, $\beta_k=\sqrt{1-4 m_k^2/s}$.

Using (\ref{st43}),(\ref{st51})-(\ref{st54}) it is simple obtain
the helicity and KS amplitudes:
\begin{eqnarray}
&&T_{hel}\left (\lambda,-\lambda,\nu_1,\nu_2\right )=
\frac{4\pi\alpha}{s}
\left( \right.
\delta_{\nu_1,-\nu_2}
\left\{  \right.
(-s) \left ( 1+\lambda \nu_1 cos\hskip 1pt \theta  \right )
\left [ Q_f+R_z \left( g_v^e -\lambda g_a^e \right )
\left ( g_v^f -\nu_1 \beta_k g_a^f \right ) \right ]
\left.  \right\}+
\nonumber \\ &&
\delta_{\nu_1,\nu_2}
\left\{\right.
2 \lambda\sqrt{s} \hskip 1pt m_k \hskip 1pt sin \hskip 1pt \theta
\left [Q_f+R_z g_v^f
\left( g_v^e -\lambda g_a^e \right)
\right ]
\left.
\right\}
\left. \right),
\label{st63}
\end{eqnarray}
\begin{eqnarray}
&&T_{KS}\left (\lambda,-\lambda,\nu_1,\nu_2\right )=
\frac{4\pi\alpha}{s} \left( \right. \sqrt{ \frac{s}{1- \beta_k^2
\hskip 1pt cos^2 \hskip 1pt \theta }} \left\{  \right.
\delta_{\nu_1,-\nu_2} (-1) \sqrt{s} \beta_k \hskip 1pt sin \hskip
1pt \theta \left (1+\lambda \nu_1 \beta_k cos\hskip 1pt \theta
\right ) \nonumber \\ && \left [ Q_f+R_z \left( g_v^e -\lambda
g_a^e \right ) \left ( g_v^f -\nu_1 g_a^f \right ) \right ]
\left.  \right\}+ \delta_{\nu_1,\nu_2} \left\{\right. 2
\left(\lambda+\nu_1\right) \hskip 1pt m_k \left [Q_f+R_z \left(
g_v^e -\lambda g_a^e \right)\right.
 \nonumber \\ &&
\left. \left ( g_v^f +\nu_1 \beta_k g_a^f \hskip 1pt cos \hskip 1pt
\theta\right ) \right ] \left. \right\} \left. \right).
\label{st64}
\end{eqnarray}
The helicity amplitude (\ref{st63}) coincide
with the amplitude, which was obtained in Ref.\cite{sp11}
up to the phase factor.

In Eqs.(\ref{st63}) and (\ref{st64}) $\nu_1,\nu_2$
correspond to the values of the helicity and KS fermion's polarization
respectively. As one can see, the helicity and KS amplitudes have the
different angular
dependence for same values $\nu_1, \nu_2$ . In Fig.2
we represent the angular cross sections
$d\sigma^{KS}_{1,-1,1,-1}\left ( z  \right )/dz$ and
$d\sigma^{hel}_{RL,RL}\left ( z  \right )/dz$ ($z=cos\hskip 1pt \theta$)
of the process $e^+e^- \to t \bar t$ with $\sqrt{s}=0.5~~TeV$.

\begin{figure}
\begin{center}
\resizebox{0.48\textwidth}{!}{
\includegraphics{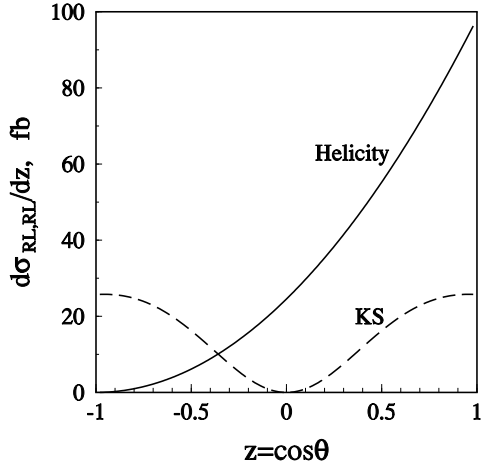}}
\end{center}
\caption{The angular cross section for the helicity and KS
polarized states}
\end{figure}

From our example we see, that it is incorrect to call the KS
states helicity ones (see,\cite{sp8}). It is to verify that the
unpolarized cross sections, which can obtained from the helicity
and KS amplitudes are coincide.

\section {Concluding Remarks}

We have presented a new spinor techniques method
for calculating the amplitudes of processes involving massive
fermions with arbitrary polarizations.
The method allows one obtain to a Feynman amplitude
in terms of spinor products.

We determined , that all possible spinor product can be
calculated with the help of the two functions only.
We have received the analytical
expressions of spinor products for vectors of polarizations
of fermions which frequently in the physical appendieces (helicity,
KS and $z$-state).  Are discussed possible unstable situations,
which can arise in the numerical calculations.

Our procedure is not more complex than CALCUL spinor
techniques for massless fermions. The method can be adopted
to both analytical and numerical computations.
The our variant of spinor
techniques is convenient for a realization on
the computer, as massless variant.
As an example, all procedure of reduction of Feynman
diagram to spinor products
are constructed with the help of the simple rule-based
program in Mathematica.

We obtained the spinor Chisholm identities for massive fermions
with arbitrary polarization.

As an illustration, the expressions are given for the amplitudes of
electron-positron annihilation into fermions-pairs for helicity
and KS-states.

\section {Acknowledgements}

I express my gratitude to the Abdus Salam ICTP for warm hospitality and
financial support. I am grateful to S.Randjbar-Daemi for the opportunity to
work at ICTP. I am grateful to I.Caprini for the help and discussions of the
results.

\section {Appendix A}

We can write the Feynman amplitudes in terms of
the current object (\ref{st60}). For example
the matrix element (\ref{st14}) can be rewritten
in terms of the dot products of the $J^{\mu}$
\begin{equation}
T_{\gamma} \left ( \lambda_1,\lambda_2;\nu_1,\nu_2 \right)=
Q_f\hskip 1pt \left ( -\nu_1 \nu_2 \right )
\left\{
J_v \left (p_2,p_1,s_{p_2},s_{p_1};\lambda_2,\lambda_1 \right ).
J_v\left (k_1,k_2,s_{k_1},s_{k_2};-\nu_1,-\nu_2 \right )
\right\}.
\label{st65}
\end{equation}
Therefore, in the calculations the currents are
important building blocks.

For massless fermions we can obtain any process in terms of
the dot products $J_v$ ,$J_u$ and the four-vectors of reaction.
This formalism ($E$-vector formalism) are used in Ref.~\cite{sp12}.
But the terms, which not reduce to the these dot products are exist,
if the all fermions are massive.

Using (\ref{st7}), the dot product $b_0 \cdot J$
reduces to the spinor products:
\begin{equation}
b_0 \cdot J_u \left( p,k,s_p ,s_k ; \lambda,\nu \right) \equiv
\overline{U}_\lambda \left( p,s_p \right)
\not\!{b}_0 U_\nu \left( k,s_k \right)=
\sum\limits_{\rho}
\left\{
\overline{U}_\lambda \left( p,s_p \right) U_\rho\left(b_0 \right)
\right\}
\left\{
\overline{U}_\rho \left(b_0 \right) U_\nu \left( k,s_k \right)
\right\}.
\label{st66}
\end{equation}
Thus we obtain analytical expressions of
the current coefficients through the components of the vectors.
For helicity states we have:
\begin{eqnarray}
&& b_0 \cdot J^{hel}_u \left( p,k,\lambda,-\lambda \right)=
\lambda
\left[ \right.
\sqrt{\left (p^0+\left|\vec {p} \right|\right )
\left (k^0-\left|\vec {k} \right|\right )}
\sqrt{\left(1+\beta_z(k)\right )\left(1-\beta_z(p)\right )}
e^{i\lambda \varphi_k}-
\nonumber \\ &&
\sqrt{\left (p^0-\left|\vec {p} \right|\right )
\left (k^0+\left|\vec {k} \right|\right )}
\sqrt{\left(1-\beta_z(k)\right )\left(1+\beta_z(p)\right )}
e^{i\lambda \varphi_p}
\left.
\right ],
\label{st67}
\end{eqnarray}
\begin{eqnarray}
&&b_0 \cdot J^{hel}_u \left( p,k,\lambda,\lambda \right)=
\sqrt{\left (p^0+\left|\vec {p} \right|\right )
\left (k^0+\left|\vec {k} \right|\right )}
\sqrt{\left(1-\beta_z(k)\right )\left(1-\beta_z(p)\right )}+
\nonumber \\ &&
\sqrt{\left (p^0-\left|\vec {p} \right|\right )
\left (k^0-\left|\vec {k} \right|\right )}
\sqrt{\left(1+\beta_z(k)\right )\left(1+\beta_z(p)\right )}
e^{i\lambda \left ( \varphi_p-\varphi_k \right)},
\label{st68}
\end{eqnarray}
\begin{eqnarray}
&&
b_3 \cdot J^{hel}_u \left( p,k,\lambda,-\lambda \right)=
\lambda
\left[ \right.
\sqrt{\left (p^0-\left|\vec {p} \right|\right )
\left (k^0+\left|\vec {k} \right|\right )}
\sqrt{\left(1+\beta_z(k)\right )\left(1-\beta_z(p)\right )}
e^{i\lambda \varphi_k}-
\nonumber \\ &&
\sqrt{\left (p^0+\left|\vec {p} \right|\right )
\left (k^0-\left|\vec {k} \right|\right )}
\sqrt{\left(1-\beta_z(k)\right )\left(1+\beta_z(p)\right )}
e^{i\lambda \varphi_p}
\left.
\right ],
\label{st69}
\end{eqnarray}
\begin{eqnarray}
&&
b_3 \cdot J^{hel}_u \left( p,k,\lambda,\lambda \right)=
\sqrt{\left (p^0-\left|\vec {p} \right|\right )
\left (k^0-\left|\vec {k} \right|\right )}
\sqrt{\left(1-\beta_z(k)\right )\left(1-\beta_z(p)\right )}+
\nonumber \\ &&
\sqrt{\left (p^0+\left|\vec {p} \right|\right )
\left (k^0+\left|\vec {k} \right|\right )}
\sqrt{\left(1+\beta_z(k)\right )\left(1+\beta_z(p)\right )}
e^{i\lambda \left ( \varphi_p-\varphi_k \right)},
\label{st70}
\end{eqnarray}
\begin{eqnarray}
&&
b_{\nu} \cdot J^{hel}_u \left( p,k,\lambda,-\lambda \right)=
\left[
\sqrt{\left (p^0-\left|\vec {p} \right|\right )
\left (k^0+\left|\vec {k} \right|\right )}-
\sqrt{\left (p^0+\left|\vec {p} \right|\right )
\left (k^0-\left|\vec {k} \right|\right )}
\right]
\nonumber \\ &&
\left\{ \right.
\delta_{\lambda,\nu}
\sqrt{\left(1-\beta_z(k)\right )\left(1-\beta_z(p)\right )}-
\sqrt{\left(1+\beta_z(k)\right )\left(1+\beta_z(p)\right )}
\nonumber \\ &&
\left(
e^{i(\varphi_k+\varphi_p)} \delta_{\lambda,1}\delta_{\nu,1}-
e^{-i(\varphi_k+\varphi_p)} \delta_{\lambda,-1}\delta_{\nu,-1}
\right)
\left.\right\},
\label{st71}
\end{eqnarray}
\begin{eqnarray}
&&
b_{\nu}\cdot J^{hel}_u \left( p,k,\lambda,\lambda \right)=
\left[
\sqrt{\left (p^0+\left|\vec {p} \right|\right )
\left (k^0+\left|\vec {k} \right|\right )}-
\sqrt{\left (p^0-\left|\vec {p} \right|\right )
\left (k^0-\left|\vec {k} \right|\right )}
\right]
\nonumber \\ &&
\left\{ \right.
\sqrt{\left(1+\beta_z(k)\right )\left(1-\beta_z(p)\right )}
\left[
\delta_{\lambda,1}\delta_{\nu,-1}\hskip 1pt e^{-i\varphi_k}+
\delta_{\lambda,-1}\delta_{\nu,1}\hskip 1pt e^{i\varphi_k}
\right]
\nonumber \\ &&
\sqrt{\left(1-\beta_z(k)\right )\left(1+\beta_z(p)\right )}
\left[
\delta_{\lambda,1}\delta_{\nu,1}\hskip 1pt e^{i\varphi_p}+
\delta_{\lambda,-1}\delta_{\nu,-1}\hskip 1pt e^{-i\varphi_p}
\right]
\left. \right\},
\label{st72}
\end{eqnarray}
where $\beta_z\left ( k \right)=k^z/\left|\vec {k} \right|$
and in Eqs.(\ref{st71})-(\ref{st72}) $\nu=\pm 1$.

\section {Appendix B}

In the center of momentum system with
the initial particles moving along $z$-axis, momenta of massive fermions
take form:
\begin{eqnarray}
&&
p_1=
\frac{\sqrt{s}}{2} (1,0,0,\beta_p)~ - ~~for~~ e^-,
\nonumber \\ &&
p_2=\frac{\sqrt{s}}{2} (1,0,0,-\beta_p)~ - ~~for~~ e^+,
\nonumber \\ &&
k_1=\frac{\sqrt{s}}{2} (1,
\beta_k \hskip 1pt sin\hskip 1 pt \theta, 0,
\beta_k \hskip 1pt cos \hskip 1pt \theta )~ - ~~for ~~f,
\nonumber \\ &&
k_2=\frac{\sqrt{s}}{2} (1,-\beta_k \hskip 1pt sin\hskip 1 pt \theta,0,
-\beta_k \hskip 1pt cos \hskip 1pt \theta)~ - ~~for~~ \bar f,
\label{st73}
\end{eqnarray}
where $\beta_p=\sqrt{1-4\hskip 1pt m_p^2/s}$.
Using (\ref{st43}) and (\ref{st51})-(\ref{st52})
we compute the helicity amplitudes :
\begin{eqnarray}
&&T_{hel}\left (\lambda,-\lambda,\nu_1,\nu_2\right )=
\frac{4\pi\alpha}{\sqrt{s}}
\left( \right.
-\delta_{\nu_1,-\nu_2}
\sqrt{s} \left (1+\lambda \nu_1 cos\hskip 1pt \theta  \right )
\nonumber \\ &&
\left [  Q_f+R_z
\left( g_v^e -\lambda\beta_{p} g_a^e \right )
\left ( g_v^f -\nu_1\beta_k g_a^f \right )
\right ] +
\delta_{\nu_1,\nu_2}
2 \lambda \hskip 1pt m_k \hskip 1pt sin \hskip 1pt \theta
\left [Q_f+R_z g_v^f
\left( g_v^e -\lambda\beta_{p} g_a^e \right)
\right ]
\left. \right),
\label{st74}
\end{eqnarray}
\begin{eqnarray}
&&
T_{hel}\left (\lambda,\lambda,\nu_1,\nu_2\right )=
\frac{4\pi\alpha}{s}
\left( \right.
4 m_k m_p  \delta_{\nu_1,\nu_2}
\left[ \right.
Q_f- R_z \hskip 1pt
\left (
g_v^e g_v^f +
\lambda \nu_1 g_a^e g_a^f \hskip 1pt
\left( s/M_Z^2 -1\right ) \hskip 1pt cos \hskip 1pt \theta
\right)
\left.
\right]
\nonumber \\ &&
-\delta_{\nu_1,-\nu_2} m_p \sqrt{s}
\hskip 1pt g_a^e R_z \hskip 1pt sin \hskip 1pt \theta
\left [ \beta_k g_v^f+ \nu_1 g_a^f
\left (4 m_k^2 /M_Z^2-1 \right)
\right]
\left. \right).
\label{st75}
\end{eqnarray}
These expressions have a compact form and in the massless limit
for $e^+$  and $e^-$ pass into (\ref{st63}).

\newpage
\section*{Figure captions}
\begin{description}
\item{\bf Fig.~1} Feynman diagrams for the process $e^+ e^-\to f \bar f$
\item{\bf Fig.~2} The angular cross section for the helicity and
KS polarized states
\end{description}
\end{document}